\input{aipcheck}

\documentclass[
    ,final            
  ]
  {aipproc}

\layoutstyle{6x9}

\begin{document}

\title{ Restoration of QCD classical symmetries in excited hadrons}


\author{ L. Ya. GLOZMAN}{
  address={Institute for physics/Theoretical physics, University of Graz,\\
Universit\"atsplatz 5, A-8010 Graz, Austria }
}

\begin{abstract}
Restoration of chiral and $U(1)_A$ symmetries in excited hadrons
is reviewed. A connection of these restorations with the semiclassical
regime in highly excited hadron is discussed.  
 A solvable confining field-theoretical toy model
 that exhibits chiral restoration is presented. Implications of
  the string description of the highly excited hadrons that suggests
 an additional dynamical symmetry of the spectra on the top of
  $U(2)_L \times U(2)_R$ are presented.
 
\end{abstract}

\maketitle


\section{Phenomenological facts, classification and definitions}
The experimental spectrum of excited hadrons, 
both baryons \cite{G1,CG,G2} and mesons \cite{G3,G4} 
suggests that the highly excited hadrons in the $u,d$ sector
fall into approximate
multiplets of $SU(2)_L \times SU(2)_R$ and $U(1)_A$ groups that
are compatible with the Poincar\'{e} invariance, 
for a short overview see 
ref. \cite{G5}.
 If 
confirmed by discovery of still missing states, this phenomenon
is referred
to as effective chiral symmetry restoration or chiral symmetry
restoration of the second kind.

It is important to precisely characterize what is implied under
effective restoration, because 
sometimes
it was (is) erroneously interpreted in the sense that the highly-excited
hadrons are in the Wigner-Weyl mode. This confusion was
a source of a controversy \cite{JPS1,CG2}, which has been overcome,
however \cite{JPS2}.
The mode of symmetry is defined
only by the properties of the vacuum. If symmetry is spontaneously
 broken in the vacuum, then it is the Nambu-Goldstone mode and the
 {\it whole} spectrum of excitations on the top of the vacuum is in
 the Nambu-Goldstone mode. However, it may happen that the role
 of the chiral symmetry breaking condensates of the vacuum becomes
 progressively irrelevant in excited states. This means that the
 chiral symmetry breaking effects (dynamics) become less and less
 important in the highly excited hadrons and asymptotically 
 the states approach the regime where their properties are determined
 by the underlying unbroken chiral symmetry (i.e. by the symmetry in
 the Wigner-Weyl mode). One of the particular consequences of the
 chiral symmetry restoration in excited hadrons is that they should
 gradually decouple from the Goldstone bosons \cite{G2,CG2,JPS2,G7,JPS3,SV}.
 A hint for such a decoupling is indeed observed phenomenologically
 since the coupling constant for $h^* \rightarrow h + \pi$ decreases
 very fast higher in the spectrum (because a decay width increases
 with the mass of the resonance much slower than the phase space
 factor).

By definition this effective chiral
 symmetry restoration means the following. All
 hadrons that are created by the given interpolator, 
 $J_\alpha$, appear
as intermediate states in the two-point correlator,

\begin{equation}
\Pi =\imath \int d^4x ~e^{\imath q x} \langle 0 | T \{ J_\alpha
(x) J^\dagger_\alpha (0) \} |0\rangle. \label{corr}
\end{equation}

\noindent  Consider two
interpolating fields $J_1(x)$ and $J_2(x)$ which are
connected by a chiral transformation (or by a $U(1)_A$
transformation), $ J_1(x) = UJ_2(x)U^\dagger. $
 Then, in the Wigner-Weyl mode, $U|0\rangle = |0\rangle,$ it
follows from (\ref{corr}) that the spectra created by the
operators $J_1(x)$ and $J_2(x)$ would be identical. We know that
in QCD one finds $U|0\rangle \neq |0\rangle.$ As a consequence the
spectra of the two operators must be in general different.
However, it happens that the noninvariance of the vacuum becomes
unimportant (irrelevant) high in the spectrum. Then the spectra of
both operators become close at large masses and asymptotically
identical.  One could say, that the valence
quarks in high-lying hadrons {\it decouple} from the quark condensate
of the  vacuum.

More precisely the effective symmetry restoration is defined to occur if
two conditions are satisfied:
(i) the states
fall into approximate multiplets of $SU(2)_L \times SU(2)_R$ 
 (and of  $U(1)_A$)
and the splittings within the multiplets ( $\Delta M = M_+ -M_-$) 
vanish at $n \rightarrow \infty$ and/or $ J \rightarrow \infty$ ;
(ii) the splitting within the multiplet is much smaller
than between the two subsequent multiplets  \cite{G3,G4,G5}.
This definition is very restrictive, because the structure
of the chiral multiplets is nontrivial and is different for
mesons with different spins. The latter is a consequence of
the requirement that the chiral multiplets must satisfy the
Poincar\'{e} invariance \cite{G4}. In particular, given the set
of the standard quantum numbers $I, J^{PC}$ the meson multiplets
of $SU(2)_L \times SU(2)_R$  are

\begin{center}
{\bf J~=~0}
\end{center}

\begin{eqnarray}
(1/2,1/2)_a  &  :  &   1,0^{-+} \longleftrightarrow 0,0^{++}  \nonumber \\
(1/2,1/2)_b  &  : &   1,0^{++} \longleftrightarrow 0,0^{-+} ,
\end{eqnarray}
\bigskip
\begin{center}
{\bf J~=~2k,~~~k=1,2,...}
\end{center}

\begin{eqnarray}
 (0,0)  & :  &   0,J^{--} \longleftrightarrow 0,J^{++}  \nonumber \\
 (1/2,1/2)_a  & : &   1,J^{-+} \longleftrightarrow 0,J^{++}  \nonumber \\
 (1/2,1/2)_b  & : &   1,J^{++} \longleftrightarrow 0,J^{-+}  \nonumber \\
 (0,1) \oplus (1,0)  & :  &   1,J^{++} \longleftrightarrow 1,J^{--} 
\end{eqnarray}

\bigskip

\begin{center}
{\bf J~=~2k-1,~~~k=1,2,...}
\end{center}

\begin{eqnarray}
 (0,0)  & :  &   0,J^{++} \longleftrightarrow 0,J^{--}  \nonumber \\
 (1/2,1/2)_a  & : &   1,J^{+-} \longleftrightarrow 0,J^{--}  \nonumber \\
 (1/2,1/2)_b  & : &   1,J^{--} \longleftrightarrow 0,J^{+-}  \nonumber \\
 (0,1) \oplus (1,0)  & :  &   1,J^{--} \longleftrightarrow 1,J^{++} 
\end{eqnarray}

The $U(1)_A$ symmetry connects the opposite parity states with
the same isospin from the distinct $(1/2,1/2)_a$  and $(1/2,1/2)_b$
multiplets of $SU(2)_L \times SU(2)_R$.

The recent data on highly excited mesons from the $\bar p p$
annihilation at LEAR \cite{BUGG1,BUGG2} do support the
$SU(2)_L \times SU(2)_R$ and $U(1)_A$ restorations as can be
seen from the high-lying $\bar n n$ mesons with $J=2$.

\begin{center}
{\bf (0,0)}\\

{$\omega_2(0,2^{--})~~~~~~~~~~~~~~~~~~~~~~~f_2(0,2^{++})$}\\
\medskip
{$1975 \pm 20   ~~~~~~~~~~~~~~~~~~~~~~~~1934 \pm 20$}\\
{$2195 \pm 30   ~~~~~~~~~~~~~~~~~~~~~~~~2240 \pm 15$}\\

\bigskip
{\bf $(1/2,1/2)_a$}\\

{$\pi_2(1,2^{-+})~~~~~~~~~~~~~~~~~~~~~~~f_2(0,2^{++})$}\\
\medskip
{$2005 \pm 15   ~~~~~~~~~~~~~~~~~~~~~~~~2001 \pm 10$}\\
{$2245 \pm 60   ~~~~~~~~~~~~~~~~~~~~~~~~2293 \pm 13$}\\

\bigskip
{\bf $(1/2,1/2)_b$}\\

{$a_2(1,2^{++})~~~~~~~~~~~~~~~~~~~~~~~ \eta_2(0,2^{-+})$}\\
\medskip
{ $2030 \pm 20  ~~~~~~~~~~~~~~~~~~~~~~~~2030 ~\pm ~?$}\\
{ $2255 \pm 20 ~~~~~~~~~~~~~~~~~~~~~~~~2267 \pm 14$}\\

\bigskip
{\bf (0,1)+(1,0)}\\

{$a_2(1,2^{++})~~~~~~~~~~~~~~~~~~~~~~~\rho_2(1,2^{--})$}\\
\medskip
{ $1950^{+30}_{-70}~~~~~~~~~~~~~~~~~~~~~~~~~~~1940 \pm 40$}\\
{ $2175 \pm 40  ~~~~~~~~~~~~~~~~~~~~~~~~2225 \pm 35$}\\
\end{center}

Note, that the chiral symmetry requires a doubling of some
of the radial and angular Regge trajectories for $J > 0$. This
is a highly nontrivial prediction of chiral symmetry. For example,
asymptotically some of the $\rho$-mesons lie on the trajectory that is
characterized by the chiral index $(0,1) \oplus (1,0)$ and have as
their chiral partners the $a_1$ mesons, while the other $\rho$-mesons
have $h_1$ mesons as their chiral partners and lie on the other independent
trajectory with the chiral index $(1/2,1/2)_b$.

If we look carefully at the data one notices that all possible
different chiral multiplets with the same $J$ are approximately
degenerate \cite{G4}. Then it means that all these states fall
into a reducible representation

\begin{equation}
 (0,1/2) \oplus (1/2,0)] \times [(0,1/2) \oplus (1/2,0)
\label{rep}
\end{equation}

\noindent
which combines all possible representations
for the system of massless quark and antiquark. Such a degeneracy
is consistent with the view of the excited hadron as a string
with massless quarks with definite chirality at the end points
of the string \cite{G2}.

There
are still some missing states in the multiplets with $J=0,1,3,4$ 
\cite{G3,G4} and it is a very important experimental task to
find them. This can be done in particular with the polarized
target formation experiment in $\bar p p$ at the NESR 
low-energy antiproton
ring at GSI, which will have similar or  better  characteristics
than LEAR.

\section{ Origins of chiral and $U(1)_A$ restorations}

An important question is a physical origin of chiral
and $U(1)_A$ restorations. If the spectrum is strictly
continuous and the function $R$ approaches a constant
value at large $s$, then the asymptotic freedom at large
space-like momenta together with a dispersion relation do allow
us to claim that the chiral symmetry is manifest in such
a spectral function, as it is observed e.g. in $e^+e^- \rightarrow jets$.
However, it is a trivial case and not what we actually need.
We have to consider  a (quasi)discrete spectrum where
the given hadron state is isolated.
The conjecture of ref. \cite{CG} was that may be the chiral restoration
is true
in the regime where the spectrum is  quasidiscrete and saturated
mainly by resonances.

One would expect that the Operator Product Expansion (OPE) 
could help us to find the correct spectrum of the high-lying hadrons. 
This is not
so, however.  This is
because the OPE is only an asymptotic expansion.
While such a kind of expansion is very useful in the space-like
region, it does not define any analytical solution which could
be continued to the time-like region.
This means that while the real (correct) spectrum of QCD must
be consistent with the OPE, there is an infinite amount of incorrect spectra
that  can also be consistent with the OPE. Then, if one wants to get
some information about the spectrum, one needs
to assume something else on the top of the OPE. Clearly a result then is
crucially dependent on these additional assumptions, for the recent
activity in this direction  see refs. \cite{OPE,SHIFMAN,GOLTERMAN}.
This implies that in order to really understand chiral symmetry restoration 
 one needs a microscopic insight and theory that would incorporate
{\it at the same time} chiral symmetry breaking and confinement.

A fundamental insight into phenomenon can be obtained from the
semiclassical expansion of the functional integral directly
in the time-like region \cite{G5}.
We know that the axial anomaly  as well as
the spontaneous breaking of chiral symmetry in QCD is an effect
of quantum fluctuations of the quark field.
The latter can
generally be seen from the definition of the quark condensate,
which is a closed quark loop.
This closed quark loop explicitly contains a factor $\hbar$. The
chiral symmetry breaking, which is necessarily a nonperturbative
effect, is actually a (nonlocal) coupling of a quark line with the
closed quark loop, which is a graphical representation of the Schwinger-Dyson (gap)
equation. 

At large $n$ (radial quantum number) or at large angular
momentum $J$ we know that in quantum systems the {\it semiclassical}
approximation  {\it must} work. Physically this approximation
applies in these cases because the de Broglie wavelength of
particles in the system is small in comparison with the
scale that characterizes the given problem. In such a system
as a hadron the scale is given by the hadron size while the
wavelength of valence quarks is given by their momenta. Once
we go high in the spectrum the size of hadrons increases as well as
 the typical momentum of valence quarks.

The semiclassical approximation  applies when the
action in the system $S \gg \hbar$.
In this case the whole amplitude (path integral) is dominated by
the classical path (stationary point) and those paths that 
are infinitesimally
close to the classical path.  In other words, in the semiclassical case the quantum
fluctuations effects are strongly suppressed and vanish asymptotically.
Then the correlation function  can be expanded
in powers of $\hbar/S$. The leading contribution 
 is a tree-level contribution to the path integral
 and keeps chiral symmetries of the classical
Lagrangian. It contains no quantum fluctuations
of the valence quark lines.
Its contribution is of the order
$(\hbar/S)^0$. The quantum fluctuations of the quark lines
as well as the vacuum fermion loops contribute at the
subleading orders in $(\hbar/S)$  and hence are
suppressed in hadrons with large intrinsic action $S$.
Then it follows 
that in a hadron with
large enough radial quantum number  $n$ or  $J$, where action is large,  
the loop
contributions must be relatively
suppressed and vanish asymptotically. Hence in
such systems both
the chiral and $U(1)_A$ symmetries should be approximately restored.
This is precisely what we see
phenomenologically.
Note that the semiclassical expansion is not an expansion in
the coupling constant, which is large in the nonperturbative regime.

\section{A solvable toy model}

While the argument presented above is general and solid enough,
 a detailed microscopical picture is missing. Then to see
how all this works one needs a solvable field-theoretical model.             
The model must be chirally
symmetric and confining.
Such a model is known \cite{Orsay,BR,L}. This model can be considered as a 
generalization
of the large $N_c$ `t Hooft model (QCD in 1+1 dimensions) \cite{HOOFT} 
to 3+1 dimensions.
In both models the only interaction between quarks is the instantaneous
infinitely raising Lorentz-vector linear potential. Then chiral symmetry
 breaking
is described by the standard   summation of the valence quarks
self-interaction loops (the Schwinger-Dyson or gap
equations), while mesons are obtained from the Bethe-Salpeter equation for the
quark-antiquark bound states.
Restoration of chiral symmetry in excited heavy-light mesons has been
previously studied with the quadratic confining potential \cite{KNR}.

\begin{figure}
\includegraphics[height=5.5cm]{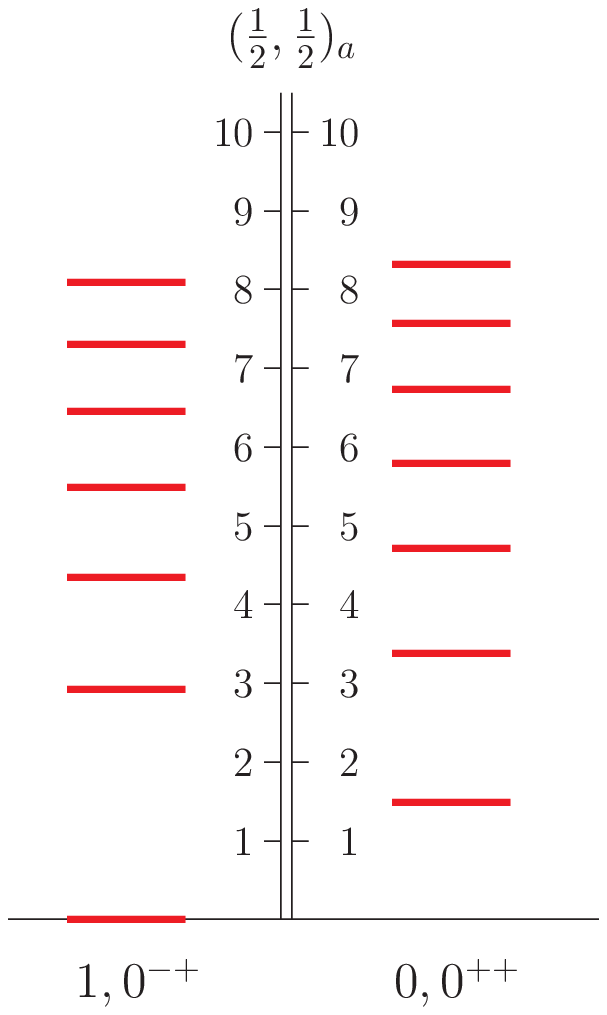}
\includegraphics[height=5.5cm,clip=]{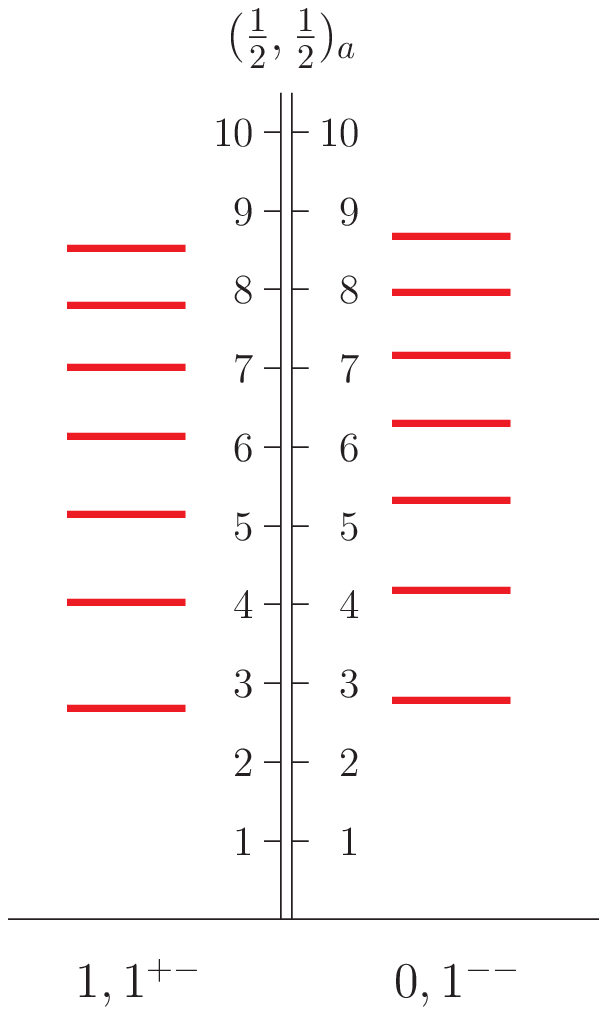}
\includegraphics[height=5.5cm,clip=]{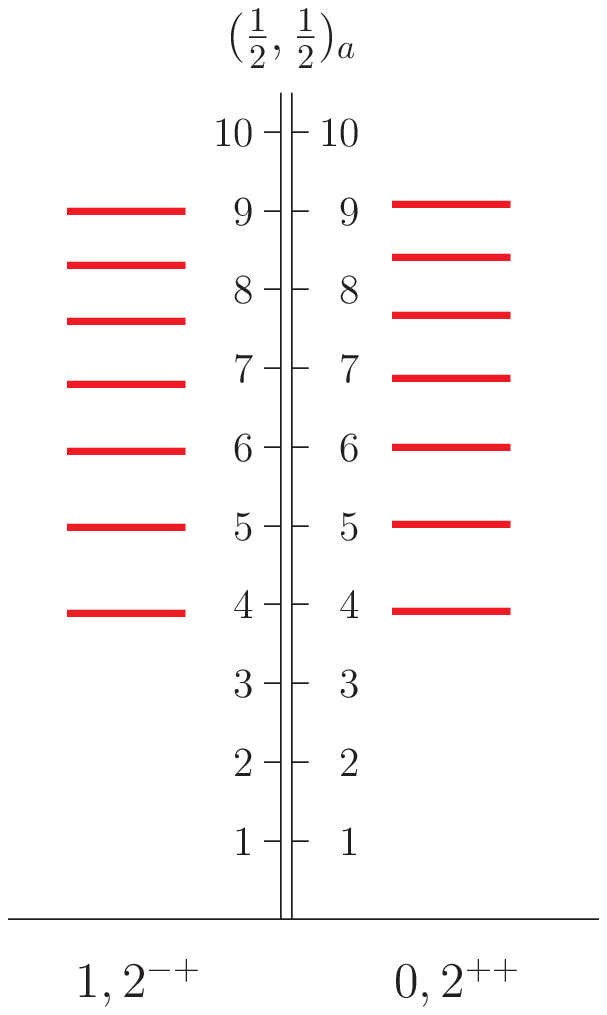}
\caption{J=0,1 and 2 spectra for mesons in $(1/2,1/2)_a$
representations.}
\end{figure}

The results for excited light-light mesons with
the linear  potential are reported in ref. \cite{WG} and presented
in Fig. 1. 
The excited states fall into
approximate chiral multiplets and
a very fast restoration of
both $SU(2)_L \times SU(2)_R$ and $U(1)_A$ symmetries with increasing of
$J$ and essentially more slow restoration with increasing of $n$ is seen.
In Fig. 2 the angular Regge trajectories are shown.
 They exhibit deviations from the linear behavior.
This fact is obviously related to the
chiral symmetry breaking effects for lower mesons.
In  the limit $n \rightarrow \infty$ and/or  $J \rightarrow \infty$ 
one observes a complete degeneracy of all multiplets, which means
that the states fall into representation (\ref{rep}).
 This means that in this limit the
quark loop effects disappear completely \cite{G5,G6}.

\begin{figure}
\includegraphics[width=0.7\hsize]{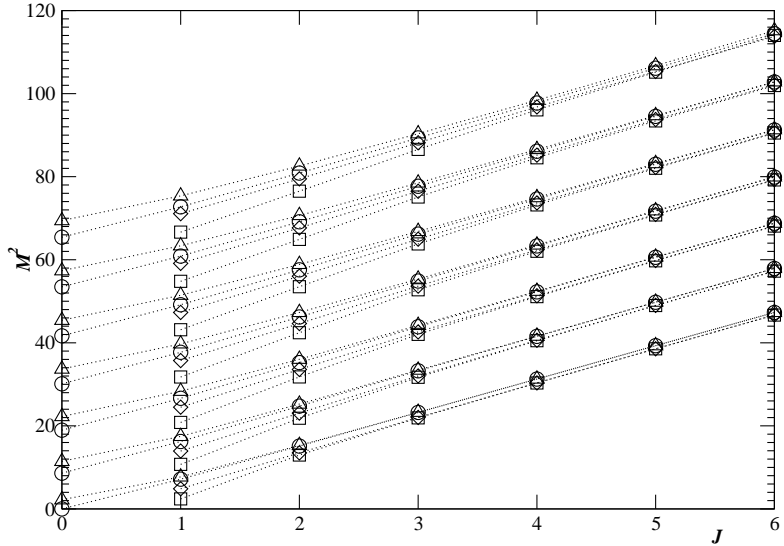}
\caption{Angular  Regge trajectories for isovector mesons
with $M^2$ in units of $\sigma$.
Mesons of the chiral multiplet $(1/2,1/2)_a$ are indicated by circles,
of $(1/2,1/2)_b$ by triangles, and of $(0,1)\oplus(1,0)$
by squares ($J^{++}$ and $J^{--}$ for even and odd $J$, respectively)
and diamonds ($J^{--}$ and $J^{++}$ for even and odd $J$, respectively).
}
\end{figure}

A few comments about physics which is behind these results. 
The chiral symmetry breaking Lorentz-scalar dynamical mass
of quarks arises via loop dressing of quarks and represents
effects of quantum fluctuations of the quark field \cite{G6}.
A key feature of this dynamical mass is that it is strongly
momentum dependent and vanishes at large quark momenta.
When one increases excitation energy of a hadron, one also
increases a typical momentum of valence quarks.
Consequently, the chiral symmetry violating Lorentz-scalar dynamical
mass of quarks
becomes small and asymptotically vanishes  in highly excited 
hadrons. Hence,
chiral and $U(1)_A$ symmetries get approximately restored \cite{G1,G6,G7}. 

Exactly the
same reason implies a decoupling of these hadrons from the 
Goldstone bosons \cite{G2,G7}.
The coupling of the Goldstone bosons to the valence
quarks is regulated via the axial current conservation by the
Goldberger-Treiman relation.
Then the coupling constant must be proportional to the Lorentz-scalar
dynamical mass of valence quarks and vanishes at larger momenta. This represents
a microscopical mechanism of decoupling which is required by the general considerations
of chiral symmetry in the Nambu-Goldstone mode \cite{CG2,JPS3,SV}.

\section{Chiral multiplets and string}

There are certain phenomenological evidences that the chiral multiplets
of excited baryons and mesons with {\it different} spins cluster at some 
energies \cite{G2,AF,SV,ICHEP}. This implies that one observes higher symmetry
that includes chiral $U(2)_L \times U(2)_R$ as a subgroup. Presumably
this additional degeneracy reflects a dynamical symmetry of the string.
Indeed, like the Veneziano amplitude, the spectrum of the open bosonic
string is degenerate with respect to the orbital momentum of the
string \cite{STRING}. A mathematical description of a hadron as a
string with quarks at the ends that have definite chirality \cite{G2}
is an open question.

\bigskip 
This work was supported by the Austrian Science Fund 
(projects P16823-N08 and P19168-N16).

\end{document}